\documentclass[12pt]{article}
  \usepackage{amsfonts}
  \usepackage{amsmath}
\usepackage{amssymb}
\usepackage{amscd}
\usepackage[dvips]{graphicx}
\begin{document}
~~
\bigskip
\bigskip
\begin{center}
{\Large {\bf{{{Generalized twist deformations of  Poincare and
Galilei Hopf algebras }}}}}
\end{center}
\bigskip
\bigskip
\bigskip
\begin{center}
{{\large ${\rm {Marcin\;Daszkiewicz}}$ }}
\end{center}
\bigskip
\begin{center}
{ {{{Institute of Theoretical Physics\\ University of Wroc{\l}aw pl.
Maxa Borna 9, 50-206 Wroc{\l}aw, Poland\\ e-mail:
marcin@ift.uni.wroc.pl}}}}
\end{center}
\bigskip
\bigskip
\bigskip
\bigskip
\bigskip
\bigskip
\bigskip
\bigskip
\begin{abstract}
The three new deformed  Poincare Hopf algebras are constructed with
use of twist procedure. The corresponding relativistic space-times
providing the sum of canonical and Lie-algebraic type of
noncommutativity are proposed. Finally, the nonrelativistic
contraction limits to the corresponding  Galilei Hopf algebras are
performed.
\end{abstract}
\bigskip
\bigskip
\bigskip
\bigskip
\bigskip
\bigskip
\bigskip
\bigskip
\bigskip
 \eject
\section{{{Introduction}}}

Due to several  theoretical arguments (see e.g. \cite{2}-\cite{2s})
the interest in  studying of space-time noncommutativity is growing
rapidly. In accordance with the Hopf-algebraic classification of all
deformations of relativistic and nonrelativistic symmetries (see
\cite{clas1}, \cite{clas2}) the most general form of space-time
noncommutativity looks as follows
\begin{equation}
[\,x_{\mu},x_\nu\,] = \theta_{\mu\nu} (x)\;,\label{nonco}
\end{equation}
 where
 \begin{equation}
 \theta_{\mu\nu} (x) = \theta_{\mu\nu}^{(0)} + \theta_{\mu\nu}^{(1)\,\rho}x_{\rho} +
 \theta_{\mu\nu}^{(2)\,\rho\tau}x_{\rho}x_{\tau}\;.\label{par}
\end{equation}

For the simplest, canonical noncommutativity $(\theta_{\mu\nu} (x) =
\theta_{\mu\nu}^{(0)})$, the corresponding Poincare Hopf algebra has
been provided in \cite{oeckl} and \cite{chi} with the use of  twist
procedure \cite{drin}-\cite{twist1}, while its nonrelativistic
counterparts have been discovered by various contraction schemes in
\cite{dasz}.

The Lie-algebraic $(\theta_{\mu\nu} (x)
=\theta_{\mu\nu}^{(1)\,\rho}x_{\rho})$ relativistic and
nonrelativistic symmetries have been proposed in \cite{kappaP} and
\cite{kappaG} respectively. In the literature they are known as
$\kappa$-Poincare and $\kappa$-Galilei Hopf algebra, which in
relativistic case
correspond to the following $\kappa$-Minkowski 
space-times
\begin{equation}
[\;{ x}_{0},{ x}_{i}\;] = \frac{i}{\kappa}{ x}_{i}\;\;\;,\;\;\;[\;{
x}_{i},{ x}_{j}\;] =0\;, \label{kappamin}
\end{equation}
with   mass-like deformation parameter $\kappa$.\\
Besides, there were proposed the twist deformations of a Lie-type at
relativistic and nonrelativistic level in \cite{lie2}, \cite{lie1}
and \cite{dasz}.

The quadratic deformation $(\theta_{\mu\nu} (x)
=\theta_{\mu\nu}^{(2)\,\rho\tau}x_{\rho}x_{\tau})$ has been studied
in \cite{paolo} and \cite{lie2}.

Unfortunately, in almost all theoretical considerations the
mentioned above quantum  space-times are considered separately.
Recently, however, there was proposed in \cite{lulya} (see also
\cite{twistkappa}) the relativistic  Hopf structure corresponding to
the so-called generalized space-time,
 with coefficients
$\theta_{\mu\nu}^{(0)}$ and $\theta_{\mu\nu}^{(1)\,\rho}$  different
than zero  simultaneously. Particulary, it has been  shown that by
canonical twist deformation of $\kappa$-Poincare  Hopf algebra, we
get $(\theta_{\mu \nu},\kappa)$-deformed symmetries associated with
the following quantum space
\begin{equation}
[\;{ x}_{0},{ x}_{i}\;] = \frac{i}{\kappa}{ x}_{i}+
i\theta_{0i}\;\;\;,\;\;\;[\;{ x}_{i},{ x}_{j}\;] =i\theta_{ij}\;.
\label{kappamin}
\end{equation}

In this article  we propose three new Poincare Hopf universal
enveloping algebras $\,{\mathcal U}_{\theta_{kl},{\kappa}}({\mathcal
P})$, $\,{\mathcal U}_{\theta_{0i},{\hat \kappa}}({\mathcal P})$ and
$\,{\mathcal U}_{\theta_{0i},{\bar \kappa}}({\mathcal P})$
corresponding to the following generalized space-times $(a,b
=1,2,3)$\footnote{The mentioned space-times are defined as the Hopf
modules of $\,{\mathcal U}_{\theta_{kl},{\kappa}}({\mathcal P})$,
$\,{\mathcal U}_{\theta_{0i},{\hat \kappa}}({\mathcal P})$ and
$\,{\mathcal U}_{\theta_{0i},{\bar \kappa}}({\mathcal P})$ Hopf
algebras respectively (see e.g. \cite{bloch}, \cite{3b},
\cite{chi}).}
\begin{equation}
[\,x_0,x_a\,] =\frac{i}{\kappa}x_i\delta_{ak}\;\;\;,\;\;\;
[\,x_a,x_b\,] =2i\theta_{kl}(\delta_{ak}\delta_{bl} -
\delta_{al}\delta_{bk}) + \frac{i}{\kappa}x_0
(\delta_{ia}\delta_{kb} - \delta_{ka}\delta_{ib})\;, \label{st1}
\end{equation}
\begin{equation}
[\,x_0,x_a\,] =\frac{i}{{\hat \kappa}}(\delta_{la}x_k -
\delta_{ka}x_l) + 2i\theta_{0i}\delta_{ia} \;\;\;,\;\;\;
[\,x_a,x_b\,] =0 \;, \label{st2}
\end{equation}
and
\begin{eqnarray}
[\,x_0,x_a\,]
=2i\theta_{0i}\delta_{ia}\;\;\;,\;\;\;
[\,x_a,x_b\,] = \frac{i}{{\bar \kappa}}\delta_{ib}(\delta_{ka}x_l -
\delta_{la}x_k) + \frac{i}{{\bar \kappa}}\delta_{ia}(\delta_{lb}x_k
- \delta_{kb}x_l) \;, \label{st3}
\end{eqnarray}
respectively. All three examples (\ref{st1})-(\ref{st3}) are
obtained by the twisting of  classical algebra with use of  the
factors defined as suitable  superposition of twist factors for
canonical and Lie-algebraic deformation of relativistic symmetry. In
other words,  used in this article algorithm follows  the procedure
\cite{drin}-\cite{twist1} used in \cite{lulya}, \cite{twistkappa},
but this time, besides twisting  the $\kappa$-Poincare algebra, we
supplement with second twist factor the  twisted  Poincare Hopf
algebras \cite{chi}, \cite{lie2}. Further, in the second step of our
investigation, we also perform three nonrelativistic contractions
(\cite{inonu}, \cite{azca1}; see also \cite{dasz}) of our
generalized Poincare Hopf structures. In such a way we get the
corresponding Galilei Hopf universal enveloping  algebras
$\,{\mathcal U}_{\xi_{kl},{\lambda}}({\mathcal G})$, $\,{\mathcal
U}_{\xi_{0i},{\hat \lambda}}({\mathcal G})$ and $\,{\mathcal
U}_{\xi_{0i},{\bar \lambda}}({\mathcal G})$ respectively.

It should be noted that there are several  motivations for  present
studies. First of all, such investigations are interesting because
they provide six new  explicit Hopf algebras. Besides, it should be
noted, that the presented algebras permit to construct the
corresponding phase-spaces (see e.g. \cite{phamelia}, \cite{lieph})
in the framework of so-called Heisenberg double procedure
\cite{twist1}. Consequently, it permits us to discuss of Heisenberg
uncertainty principle associated with such generalized quantum
space-times. Finally,  one can consider corresponding classical and
quantum relativistic and the nonrelativistic particle models. Such a
construction has been already presented  in the case of classical
nonrelativistic particle moving in  external constant force
\cite{daszwal}, \cite{daszwal1}, and the studies of  deformations
(\ref{st1})-(\ref{st3}) in a context of dynamical considerations are
postponed for further investigations.

The paper is organized as follows. In  second Section we recall
 necessary facts concerning  twist-deformed Poincare Hopf algebras
 \cite{chi},
 \cite{lie2}. In Section 3 we present    three new Poincare
Hopf structures; the corresponding generalized space-times and the
proper nonrelativistic contractions to Galilei algebras are
presented in Sections 4 and 5 respectively. The results are
summarized and discussed in the last Section.

\section{{{Twisted Poincare Hopf algebras}}}

Let us recall five canonically and Lie-algebraically twisted
Poincare Hopf algebras $\,{\mathcal U}_{\cdot}({\mathcal P})$
proposed  in \cite{chi} and \cite{lie2} respectively. All of them
are described by classical (undeformed) algebraic sector
$(\eta_{\mu\nu} = (-,+,+,+))$
\begin{eqnarray}
&&\left[ M_{\mu \nu },M_{\rho \sigma }\right] =i\left( \eta _{\mu
\sigma }\,M_{\nu \rho }-\eta _{\nu \sigma }\,M_{\mu \rho }+\eta
_{\nu \rho }M_{\mu
\sigma }-\eta _{\mu \rho }M_{\nu \sigma }\right) \;,  \notag \\
&&\left[ M_{\mu \nu },P_{\rho }\right] =i\left( \eta _{\nu \rho
}\,P_{\mu }-\eta _{\mu \rho }\,P_{\nu }\right) \;\;\;,\;\;\; \left[
P_{\mu },P_{\nu }\right] =0\;,  \label{nnn}
\end{eqnarray}
and twisted coalgebraic part
\begin{equation}
\Delta _{0}(a) \to \Delta _{\cdot }(a) = \mathcal{F}_{\cdot }\circ
\,\Delta _{0}(a)\,\circ \mathcal{F}_{\cdot
}^{-1}\;\;\;,\;\;\;
S_{\cdot}(a) =u_{\cdot }\,S_{0}(a)\,u^{-1}_{\cdot }\;,\label{fs}
\end{equation}
with $\Delta _{0}(a) = a \otimes 1 + 1 \otimes a$, $S_0(a) = -a$ and
$u_{\cdot }=\sum f_{(1)}S_0(f_{(2)})$ (we use Sweedler's notation
$\mathcal{F}_{\cdot }=\sum f_{(1)}\otimes f_{(2)}$).\\
Present in the above formula twist element $\mathcal{F}_{\cdot } \in
{\mathcal U}_{\cdot}({\mathcal P}) \otimes {\mathcal
U}_{\cdot}({\mathcal P})$ satisfies the classical cocycle  condition
\cite{twist}, \cite{twist1}
\begin{equation}
{\mathcal F}_{{\cdot }12} \cdot(\Delta_{0} \otimes 1) ~{\cal
F}_{\cdot } = {\mathcal F}_{{\cdot }23} \cdot(1\otimes \Delta_{0})
~{\mathcal F}_{{\cdot }}\;, \label{cocyclef}
\end{equation}
and the normalization condition
\begin{equation}
(\epsilon \otimes 1)~{\cal F}_{{\cdot }} = (1 \otimes
\epsilon)~{\cal F}_{{\cdot }} = 1\;, \label{normalizationhh}
\end{equation}
with ${\cal F}_{{\cdot }12} = {\cal F}_{{\cdot }}\otimes 1$ and
${\cal F}_{{\cdot }23} = 1 \otimes {\cal F}_{{\cdot }}$\footnote{All
carriers of considered twist factors are Abelian.}.

In the case of first, canonically deformed algebra ${\mathcal
U}_{\theta_{kl}}({\mathcal P})$, the twist element looks as follows
$(\theta_{kl} = -\theta_{lk})$
\begin{equation}
{\mathcal{F}}_{\theta_{kl} }=\exp i\left(\theta_{kl}P_{k }\wedge
P_{l}  \right)\;\;\;;\;\;\;[\;k,l - {\rm fixed}\;]\;\;\;,\;\;\; a
\wedge b = a \otimes b -b \otimes a \label{gfactor}
\end{equation}
and, in accordance with (\ref{fs}), the corresponding coproduct
sector takes the form
\begin{eqnarray}
\Delta_{\theta_{kl}}(P_\mu)&=&\Delta_0(P_\mu)\;, \label{dlww3v}\\
\Delta _{\theta_{kl} }(M_{\mu \nu }) &=&\Delta _{0}(M_{\mu \nu })-
\theta_{kl}[(\eta _{k \mu }P_{\nu }-\eta _{k \nu }\,P_{\mu })\otimes
P_{l }+P_{k}\otimes (\eta_{l
\mu}P_{\nu}-\eta_{l \nu}P_{\mu})]\nonumber\\
&+& \theta_{kl}[(\eta _{l \mu }P_{\nu }-\eta _{l \nu }\,P_{\mu
})\otimes P_{k }+P_{l}\otimes (\eta_{k \mu}P_{\nu}-\eta_{k
\nu}P_{\mu})]\;.\label{zadruzny}
\end{eqnarray}
The antipodes and counits remain undeformed
\begin{eqnarray}
S_{0}(P_{\mu }) =-P_{\mu }\;\;\;\;,\;\;\;\; S_{0}(M_{\mu \nu })
=-M_{\mu \nu }\;\;\;\;,\;\;\;\;
 \epsilon(M_{\mu \nu })=\epsilon(P_\mu)=0\;.  \label{zadruga1}
 \vspace{0.3cm}
\end{eqnarray}

For the second considered algebra ${\mathcal
U}_{\theta_{0i}}({\mathcal P})$ we have the following twist element
\begin{equation}
{\mathcal{F}}_{\theta_{0i}}=\exp i\left({{\theta}_{0i}}P_{0 }\wedge
P_{i}  \right)\;\;\;;\;\;\;[\;i - {\rm
fixed}\;]\;\;\;,\;\;\;(\theta_{0i} = -\theta_{i0})\;,
\label{fggfactor}
\end{equation}
and the corresponding coproduct sector is given by
\begin{eqnarray}
\Delta_{\theta_{0i}}(P_\mu)&=&\Delta_0(P_\mu)\;, \label{zcoppy1}\\
\Delta _{\theta_{0i} }(M_{\mu \nu })&=&\Delta _{0}(M_{\mu \nu })-
\theta_{0i}[(\eta _{0 \mu }P_{\nu }-\eta _{0 \nu }\,P_{\mu })\otimes
P_{i }+P_{0}\otimes (\eta_{i
\mu}P_{\nu}-\eta_{i \nu}P_{\mu})]\nonumber\\
&+& \theta_{0i}[(\eta _{i \mu }P_{\nu }-\eta _{i \nu }\,P_{\mu
})\otimes P_{0 }+P_{i}\otimes (\eta_{0 \mu}P_{\nu}-\eta_{0
\nu}P_{\mu})]\;.\label{zcoppy100}
\end{eqnarray}
The antipodes and counits become classical.

In the case of Lie-algebraically deformed Hopf algebra  ${\mathcal
U}_{\kappa}({\mathcal P})$ the twist factor and coproducts look as
follows
\begin{equation}
{\mathcal{F}}_{\kappa }=\exp \frac{i}{2\kappa }\left(P_k \wedge
M_{i0}
 \right)\;\;\;\; [\;i,k - {\rm fixed},\;\;i \neq k\;]\;,
\label{dghfactor}
\end{equation}
\begin{eqnarray}
\Delta_{{\kappa}}(P_\mu)&=&\Delta
_0(P_\mu)+\sinh(\frac{1}{2{\kappa}} P_k )\wedge
\left(\eta_{i \mu}P_0 -\eta_{0 \mu}P_i \right)\nonumber\\
&+&(\cosh(\frac{1}{2{\kappa}}  P_k )-1)\perp \left(\eta_{i \mu}P_i
-\eta_{0 \mu}P_0 \right)\label{coppy1}\;,
\end{eqnarray}
\begin{eqnarray}
\Delta_{ {\kappa}}(M_{\mu\nu})&=&\Delta_0(M_{\mu\nu})+\frac{1}{2
{\kappa}}M_{i 0 }\wedge  \left(\eta_{\mu
k }P_\nu-\eta_{\nu k}P_\mu\right)\notag\\
&+&i\left[M_{\mu\nu},M_{i 0 }\right]\wedge
\sinh(\frac{1}{2 {\kappa}} P_k ) \notag \\
&-&\left[\left[%
M_{\mu\nu},M_{i 0 }\right],M_{i 0 }\right]\perp
(\cosh(\frac{1}{2 {\kappa}}  P_k  )-1)  \label{coppy100} \\
&+&\frac{1}{2 {\kappa}}M_{i 0 }\sinh(\frac{1}{2 {\kappa}} P_k )\perp
 \left(\psi_k P_i -\chi_k P_0 \right) \nonumber \\
&-&\frac{1}{2 {\kappa}} \left(\psi_k P_0 -\chi_k P_i \right)\wedge
M_{i 0 }(\cosh(\frac{1}{2 {\kappa}} P_k )-1) \;,\notag
\end{eqnarray}
\\
where
\begin{equation}
\psi_k =\delta_{\nu k }\delta_{0 \mu}-\delta_{\mu k }\delta_{0
\nu}\;\;\;,\;\;\; \chi_k =\delta_{\nu k }\delta_{i \mu}-\delta_{\mu
k }\delta_{i \nu}\;\;\;,\;\;\;a \perp b = a \otimes b + b\otimes
a\;. \label{dodatek1}
\end{equation}
The antipodes and coproducts remain undeformed.

In the case of two remaining Lie-algebraic Hopf structures  the
twist factors are given by
\begin{eqnarray}
&&{\mathcal{F}}_{\hat{\kappa}}=\exp \frac{i}{2\hat{\kappa}}\left(P_0
\wedge {M_{kl}} \right) \;\;\;\;
[\;k,l - {\rm fixed}\;]\;, \label{czartwor500}\\
&&{\mathcal{F}}_{\bar{\kappa}}=\exp \frac{i}{2
\bar{\kappa}}\left(P_i \wedge {M_{kl}} \right) \;\;\;\; [\;i,k,l -
{\rm fixed},\;\;i \neq k,l\;]\;, \label{czartwor600}
\end{eqnarray}
while the corresponding coproducts take the form
\begin{eqnarray}
\Delta_{\hat{\kappa}}(P_\mu)&=&\Delta
_0(P_\mu)+\sin(\frac{1}{2\hat{\kappa}} P_{0} )\wedge
\left(\eta_{k \mu}P_l -\eta_{l \mu}P_k \right)\nonumber\\
&+&(\cos(\frac{1}{2\hat{\kappa}}  P_{0} )-1)\perp \left(\eta_{k
\mu}P_k +\eta_{l \mu}P_l \right)\;, \label{czartworfff}
\end{eqnarray}
\begin{eqnarray}
\Delta_{\hat{\kappa}}(M_{\mu\nu})&=&\Delta_0(M_{\mu\nu})+\frac{1}{2\hat{\kappa}}M_{k
l }\wedge  \left(\eta_{\mu
0 }P_\nu-\eta_{\nu 0 }P_\mu\right)\notag\\
&+&i\left[M_{\mu\nu},M_{k l }\right]\wedge
\sin(\frac{1}{2\hat{\kappa}}P_{0} ) \notag \\
&+&\left[\left[%
M_{\mu\nu},M_{k l }\right],M_{k l }\right]\perp
(\cos(\frac{1}{2\hat{\kappa}} P_{0} )-1)  \nonumber \\
&+&\frac{1}{2\hat\kappa}M_{k l }\sin(\frac{1}{2\hat{\kappa}} P_{0}
)\perp
 \left(\psi_{0} P_k -\chi_{0} P_l \right) \label{czartworcopp2} \\
&+&\frac{1}{2\hat\kappa} \left(\psi_{0} P_l +\chi_{0} P_k
\right)\wedge M_{k l }(\cos(\frac{1}{2\hat{\kappa}} P_{0} )-1)
\;,\notag
\end{eqnarray}
and
\begin{eqnarray}
\Delta_{\bar\kappa}(P_\mu)&=&\Delta
_0(P_\mu)+\sin(\frac{1}{2\bar\kappa} P_{i} )\wedge
\left(\eta_{k \mu}P_l -\eta_{l \mu}P_k \right)\nonumber\\
&+&(\cos(\frac{1}{2\bar\kappa}  P_{i} )-1)\perp \left(\eta_{k
\mu}P_k +\eta_{l \mu}P_l \right)\;, \label{nowageneracja1}
\end{eqnarray}
\begin{eqnarray}
\Delta_{\bar\kappa}(M_{\mu\nu})&=&\Delta_0(M_{\mu\nu})+\frac{1}{2\bar\kappa}M_{k
l }\wedge  \left(\eta_{\mu
i }P_\nu-\eta_{\nu i }P_\mu\right)\notag\\
&+&i\left[M_{\mu\nu},M_{k l }\right]\wedge
\sin(\frac{1}{2\bar\kappa}P_{i} ) \notag \\
&+&\left[\left[%
M_{\mu\nu},M_{k l }\right],M_{k l }\right]\perp
(\cos(\frac{1}{2\bar\kappa} P_{i} )-1)  \nonumber \\
&+&\frac{1}{2\bar\kappa}M_{k l }\sin(\frac{1}{2\bar\kappa} P_{i}
)\perp
 \left(\psi_{i} P_k -\chi_{i} P_l \right) \label{nowageneracja3} \\
&+&\frac{1}{2\bar\kappa} \left(\psi_{i} P_l +\chi_{i} P_k
\right)\wedge M_{k l }(\cos(\frac{1}{2\bar\kappa} P_{i} )-1)
\;,\notag
\end{eqnarray}
respectively, with
\begin{equation}
\psi_\lambda =\eta_{\nu \lambda }\eta_{l \mu}-\eta_{\mu \lambda
}\eta_{l \nu}\;\;\;,\;\;\; \chi_\lambda =\eta_{\nu \lambda }\eta_{k
\mu}-\eta_{\mu \lambda }\eta_{k \nu}\;.
\end{equation}
The antipodes and counit remain classical.

It should be noted, that all above algebras can be derived from
relativistic  classical r-matrices $r_{\cdot } \in {\mathcal
U}_{\cdot}({\mathcal P}) \otimes {\mathcal U}_{\cdot}({\mathcal
P})$, which are given by
\begin{equation}
r_{\theta_{kl}}=\theta_{kl}P_{k}\wedge P_{l} \;\;\;\;[\;k,l - {\rm
fixed}\;]\;, \label{grmatix1}
\end{equation}
\begin{equation}
r_{\theta_{0i}}={{\theta}_{0i}}P_{0 }\wedge P_{i} \;\;\;\;[\;i -
{\rm fixed}\;]\;, \label{grmatix2}
\end{equation}
\begin{equation}
\;\;\;r_{{\kappa}}=\frac{1}{2{\kappa}}P_{k}\wedge M_{i0} \;\;\;\;\;
[\;i,k - {\rm fixed},\;\;i \neq k\;]\;, \label{grmatix3}
\end{equation}
\begin{equation}
r_{{\hat \kappa}}=\frac{1}{2{\hat \kappa}}P_{0}\wedge M_{kl}\;\;\;\;
[\;k,l - {\rm fixed}\;]\;, \;\;\;\;\label{grmatix4}
\end{equation}
and
\begin{equation}
r_{{\bar \kappa}}=\frac{1}{2{\bar \kappa}}P_{i}\wedge M_{kl}\;\;\;
[\;i,k,l - {\rm fixed},\;\;i \neq k,l\;]\;. \label{grmatix5}
\end{equation}
The matrices (\ref{grmatix1})-(\ref{grmatix5}) satisfy the classical
Yang-Baxter equation (CYBE)
\begin{equation}
[[\;r_{\cdot},r_{\cdot}\;]] = [\;r_{\cdot 12},r_{\cdot13} + r_{\cdot
23}\;] + [\;r_{\cdot 13}, r_{\cdot 23}\;] = 0\;, \label{cybe}
\end{equation}
where  the  symbol $[[\;\cdot,\cdot\;]]$ denotes the Schouten
bracket and $r_{\cdot 12} =r_{\cdot} \wedge 1$, $r_{\cdot 13} =
r_{\cdot 1} \wedge 1\wedge r_{\cdot 2}$, $r_{\cdot 23} = 1\wedge
r_{\cdot}$, $r_{\cdot } = r_{\cdot 1} \wedge r_{\cdot 2}$.

Obviously, for parameters $\theta_{kl}$, $\theta_{0i}$ running to
zero and parameters $\kappa$, ${\hat \kappa}$, ${\bar \kappa}$
approaching infinity all  above algebras become classical.

\section{{{Generalized twisted Poincare Hopf algebras}}}

In this Section we introduced the generalized twisted algebras
described by  proper sums of matrices
(\ref{grmatix1})-(\ref{grmatix5}).

\subsection*{{\bf i)} relativistic
$\bf(\theta_{kl},\kappa)$-deformation}

$~~~~~$Let us start with the following classical r-matrix
\begin{eqnarray}
   r_{\theta_{kl},{\kappa}}  = \frac{1}{2{\kappa}}P_k
\wedge M_{i0} + \theta_{kl}P_{k}\wedge P_{l}  \;,
\label{rge1}
\end{eqnarray}
defined as the sum of r-matrices (\ref{grmatix1}) and
(\ref{grmatix3}) with  indices $k$, $l$ different than
$i$\footnote{The carrier of  matrix (\ref{rge1}) is Abelian.}. One
can check that it satisfies the CYBE equation (\ref{cybe}).

 We find the corresponding deformed coproduct sector in
two steps. Firstly, we twist classical Poincare algebra $\,{\mathcal
U}_{0}({\mathcal P})$ with use of the factor (\ref{gfactor}) or
(\ref{dghfactor}). In such a way we get the Hopf algebra
$\,{\mathcal U}_{\theta_{kl}}({\mathcal P})$ or $\,{\mathcal
U}_{\kappa}({\mathcal P})$ described in pervious Section. Next,
following \cite{lulya} and \cite{twistkappa}, we twist the
coalgebraic sector of $\,{\mathcal U}_{\theta_{kl}}({\mathcal P})$
or $\,{\mathcal U}_{\kappa}({\mathcal P})$ with the use of  twist
factor\footnote{The considered (second) factor corresponds to the
matrix (\ref{grmatix3}) or (\ref{grmatix1}) respectively.}
satisfying $\theta_{kl}$- or $\kappa$-deformed cocycle  condition
(see formulas (\ref{dlww3v}), (\ref{zadruzny}) or (\ref{coppy1}),
(\ref{coppy100}))
\begin{equation}
{\bar{\mathcal F}}_{{\kappa/\theta_{kl}}12}
\cdot(\Delta_{\theta_{kl}/\kappa} \otimes 1) ~{\bar{\cal
F}_{\kappa/\theta_{kl} }} = {\bar{\mathcal
F}}_{{\kappa/\theta_{kl}}23} \cdot(1\otimes
\Delta_{\theta_{kl}/\kappa}) ~{\bar{\mathcal
F}}_{{\kappa/\theta_{kl}}}\;, \label{cocyclefnext}
\end{equation}
and normalization condition
\begin{equation}
(\epsilon \otimes 1)~{\bar{\cal F}}_{{\kappa/\theta_{kl}}} = (1
\otimes \epsilon)~{\bar{\cal F}}_{{\kappa/\theta_{kl}}} = 1\;.
\label{normalizationhhnext}
\end{equation}
One can check that the solutions ${\bar{\mathcal
F}}_{{\kappa/\theta_{kl}}}$ of the equations (\ref{cocyclefnext}),
(\ref{normalizationhhnext}) are the same as for classical coproduct
$\Delta_0(a)$, i.e. they are given by the formulas (\ref{dghfactor})
or (\ref{gfactor}) respectively\footnote{Consequently, in accordance
with the  Abelian character   of  matrix (\ref{rge1}), the total
twist factor takes the form ${{\mathcal F}}_{{\theta_{kl},\kappa}} =
{{\mathcal F}}_{{\kappa}}\cdot {{\mathcal F}}_{{\theta_{kl}}} =
{{\mathcal F}}_{{\theta_{kl}}}\cdot {{\mathcal F}}_{{\kappa}} = {\rm
exp}\,i(\frac{1}{2{\kappa}}P_k \wedge M_{i0} +
\theta_{kl}P_{k}\wedge P_{l})$.}. In such a way we get the following
coalgebraic sector
\begin{eqnarray}
\Delta_{\theta_{kl},{\kappa}}(P_\mu)&=&\Delta
_0(P_\mu)+\sinh(\frac{1}{2 {\kappa}} P_k )\wedge
\left(\eta_{i \mu}P_0 -\eta_{0 \mu}P_i \right)\label{coa1}\\
&+&(\cosh(\frac{1}{2 {\kappa}}  P_k )-1)\perp \left(\eta_{i \mu}P_i
-\eta_{0 \mu}P_0 \right)\;,\notag
\end{eqnarray}
\begin{eqnarray}
\Delta_{\theta_{kl},{\kappa}}(M_{\mu\nu})&=&\Delta_0(M_{\mu\nu})+\frac{1}{2
{\kappa}}M_{i 0 }\wedge  \left(\eta_{\mu
k }P_\nu-\eta_{\nu k}P_\mu\right)\nonumber\\
&+&i\left[M_{\mu\nu},M_{i 0 }\right]\wedge
\sinh(\frac{1}{2 {\kappa}} P_k ) \notag \\
&-&\left[\left[%
M_{\mu\nu},M_{i 0 }\right],M_{i 0 }\right]\perp
(\cosh(\frac{1}{2 {\kappa}}  P_k  )-1) \nonumber \\
&+&\frac{1}{2 {\kappa}}M_{i 0 }\sinh(\frac{1}{2 {\kappa}} P_k )\perp
 \left(\psi_k P_i -\chi_k P_0 \right) \label{coa100} \\
&-&\frac{1}{2 {\kappa}} \left(\psi_k P_0 -\chi_k P_i \right)\wedge
M_{i 0 }(\cosh(\frac{1}{2 {\kappa}} P_k )-1) \notag
\end{eqnarray}
\begin{eqnarray}
&-& \theta ^{k l }[(\eta _{k \mu }P_{\nu }-\eta _{k \nu }\,P_{\mu
})\otimes P_{l }+P_{k}\otimes (\eta_{l
\mu}P_{\nu}-\eta_{l \nu}P_{\mu})]\nonumber\\
&+& \theta_{kl}[(\eta _{l \mu }P_{\nu }-\eta _{l \nu }\,P_{\mu
})\otimes P_{k }+P_{l}\otimes (\eta_{k \mu}P_{\nu}-\eta_{k
\nu}P_{\mu})]\nonumber\\
&~~&\nonumber\\
&+& \theta_{kl}\left[\left[M_{\mu\nu},M_{i 0 }\right],P_k\right]
\perp
\sinh(\frac{1}{2 {\kappa}} P_k )P_l\notag\\
&-& \theta_{kl}\left[\left[M_{\mu\nu},M_{i 0 }\right],P_l\right]
\perp \sinh(\frac{1}{2 {\kappa}} P_k )P_k
\notag\\
&+&i\theta_{kl}\left[\left[\left[%
M_{\mu\nu},M_{i 0 }\right],M_{i 0 }\right],P_k\right] \wedge
(\cosh(\frac{1}{2 {\kappa}}  P_k  )-1)P_l
\notag\\
&-& i\theta_{kl}\left[\left[\left[%
M_{\mu\nu},M_{i 0 }\right],M_{i 0 }\right],P_l\right] \wedge
(\cosh(\frac{1}{2 {\kappa}}  P_k  )-1)P_k\;,\notag
\end{eqnarray}
which together with  algebraic relations (\ref{nnn}) and
antipodes/counits (\ref{zadruga1})  defines the generalized Poincare
Hopf  algebra $\,{\mathcal U}_{\theta_{kl},\kappa}({\mathcal P})$.
It should be noted that for parameter $\theta_{kl}$ approaching zero
we get $\,{\mathcal U}_{\kappa}({\mathcal P})$ algebra, while for
$\kappa$ running to infinity we obtain $\,{\mathcal
U}_{\theta_{kl}}({\mathcal P})$ Poincare Hopf structure  provided in
pervious Section.

\subsection*{{\bf ii)} relativistic
$\bf(\theta_{0i},\hat{\kappa})$-deformation}

$~~~~~$Let us now turn to the second (generalized) classical
r-matrix
\begin{eqnarray}
   r_{\theta_{0i},\hat{\kappa}}  =
\frac{1}{2\hat{\kappa}}P_0 \wedge M_{kl} + {{\theta}_{0i}}P_{0
}\wedge P_{i}  \;,
\label{rge2}
\end{eqnarray}
where index $i$ is  different than $k$, $l$, 0\footnote{The carrier
of matrix (\ref{rge2}) remains Abelian.}. We see, that the above
r-matrix is defined as  a sum of r-matrices for $\,{\mathcal
U}_{{\hat \kappa}}({\mathcal P})$  and $\,{\mathcal
U}_{\theta_{0i}}({\mathcal P})$ algebras. Obviously, it satisfies
CYBE equation (\ref{cybe}).

 In order to find the corresponding Hopf
algebra we use the same prescription as in the case of constructed
above Hopf structure $\,{\mathcal U}_{\theta_{kl},\kappa}({\mathcal
P})$. It should be noted however, that this time,  in the second
step of used algorithm we twist the algebras $\,{\mathcal
U}_{{\hat\kappa}}({\mathcal P})$ or $\,{\mathcal
U}_{\theta_{0i}}({\mathcal P})$ respectively. One can check that as
before, the corresponding twist factors are the same as
(\ref{fggfactor}) or (\ref{czartwor500}), and they solve the
modified cocycle condition with respect to coproducts
(\ref{czartworfff}), (\ref{czartworcopp2}) or (\ref{zcoppy1}),
(\ref{zcoppy100})\footnote{Hence, the total twist factor can be
written as ${{\mathcal F}}_{{\theta_{0i},\hat{\kappa}}} =  {\rm
exp}\,i(\frac{1}{2\hat{\kappa}}P_0 \wedge M_{kl} +
{{\theta}_{0i}}P_{0 }\wedge P_{i})$.}. Consequently, we get
\begin{eqnarray}
\Delta_{\theta_{0i},\hat{\kappa}}(P_\mu)&=&\Delta
_0(P_\mu)+\sin(\frac{1}{2\hat{\kappa}} P_{0} )\wedge
\left(\eta_{k \mu}P_l -\eta_{l \mu}P_k \right)\label{nextczartworfff}\\
&+&(\cos(\frac{1}{2\hat{\kappa}}  P_{0} )-1)\perp \left(\eta_{k
\mu}P_k +\eta_{l \mu}P_l \right)\;,  \notag
\end{eqnarray}
\begin{eqnarray}
\Delta_{\theta_{0i},\hat{\kappa}}(M_{\mu\nu})&=&\Delta_0(M_{\mu\nu})+\frac{1}{2\hat{\kappa}}M_{k
l }\wedge  \left(\eta_{\mu
0 }P_\nu-\eta_{\nu 0 }P_\mu\right)\notag\\
&+&i\left[M_{\mu\nu},M_{k l }\right]\wedge
\sin(\frac{1}{2\hat{\kappa}}P_{0} ) \notag \\
&+&\left[\left[%
M_{\mu\nu},M_{k l }\right],M_{k l }\right]\perp
(\cos(\frac{1}{2\hat{\kappa}} P_{0} )-1)  \label{nextczartworcopp2} \\
&+&\frac{1}{2 {\hat\kappa}}M_{k l }\sin(\frac{1}{2\hat{\kappa}}
P_{0} )\perp
 \left(\psi_{0} P_k -\chi_{0} P_l \right) \nonumber \\
&+&\frac{1}{2\hat{\kappa}} \left(\psi_{0} P_l +\chi_{0} P_k
\right)\wedge M_{k l }(\cos(\frac{1}{2\hat{\kappa}} P_{0}
)-1)\nonumber\\
&~~&\nonumber\\
&-& \theta_{0i}[(\eta _{0 \mu }P_{\nu }-\eta _{0 \nu }\,P_{\mu
})\otimes P_{i }+P_{0}\otimes (\eta_{i
\mu}P_{\nu}-\eta_{i \nu}P_{\mu})]\nonumber\\
&+& \theta_{0i}[(\eta _{i \mu }P_{\nu }-\eta _{i \nu }\,P_{\mu
})\otimes P_{0 }+P_{i}\otimes (\eta_{0 \mu}P_{\nu}-\eta_{0
\nu}P_{\mu})]\nonumber\\
&+& \theta_{0i}\left[\left[M_{\mu\nu},M_{kl }\right],P_0\right]
\perp
\sin(\frac{1}{2 {\hat\kappa}} P_0 )P_i\notag\\
&-& \theta_{0i}\left[\left[M_{\mu\nu},M_{kl }\right],P_i\right]
\perp \sin(\frac{1}{2 {\hat\kappa}} P_0 )P_0
\notag\\
&-&i\theta_{0i}\left[\left[\left[%
M_{\mu\nu},M_{kl }\right],M_{kl }\right],P_0\right] \wedge
(\cos(\frac{1}{2 {\hat\kappa}}  P_0  )-1)P_i
\notag\\
&+& i\theta_{0i}\left[\left[\left[%
M_{\mu\nu},M_{kl }\right],M_{kl }\right],P_i\right] \wedge
(\cos(\frac{1}{2 {\hat\kappa}}  P_0 )-1)P_0\;.\notag
\end{eqnarray}
The above relations together with algebraic sector (\ref{nnn}) and
antipodes/counits (\ref{zadruga1}) define the generalized Poincare
Hopf algebra $\,{\mathcal U}_{\theta_{0i},{\hat \kappa}}({\mathcal
P})$. Of course, for parameter $\theta_{0i}$ running to zero we
obtain Hopf structure $\,{\mathcal U}_{{\hat \kappa}}({\mathcal
P})$, while for parameter ${\hat \kappa}$ approaching infinity we
get $\,{\mathcal U}_{\theta_{0i}}({\mathcal P})$ algebra described
in Section 2.

\subsection*{{\bf iii)} relativistic
$\bf(\theta_{0i},\bar{\kappa})$-deformation}

$~~~~~$Let us now consider the last generalized r-matrix
\begin{eqnarray}
   r_{\theta_{0i},\bar{\kappa}}  =
\frac{1}{2\bar{\kappa}}P_i \wedge M_{kl} + {{\theta}_{0i}}P_{0
}\wedge P_{i}  \;,
\label{rge3}
\end{eqnarray}
defined as a sum of r-matrices for $\,{\mathcal
U}_{{\bar\kappa}}({\mathcal P})$ and $\,{\mathcal
U}_{{\theta_{0i}}}({\mathcal P})$ Hopf algebras with  index $i$
different than $k$, $l$ and $0$. It satisfies the classical
Yang-Baxter equation (\ref{cybe}).

We get the corresponding Hopf algebra by twist procedure of  the
coproducts (\ref{zcoppy1}), (\ref{zcoppy100}) or
(\ref{nowageneracja1}), (\ref{nowageneracja3}), where the proper
twist factors are exactly the same as (\ref{czartwor600}) or
(\ref{fggfactor}), and satisfy
 the following modified cocycle  condition
\begin{equation}
{\bar{\mathcal F}}_{{\bar{\kappa}/\theta_{0i}}12}
\cdot(\Delta_{\theta_{0i}/\bar{\kappa}} \otimes 1) ~{\bar{\cal
F}_{\bar{\kappa}/\theta_{0i} }} = {\bar{\mathcal
F}}_{{\bar{\kappa}/\theta_{0i}}23} \cdot(1\otimes
\Delta_{\theta_{0i}/\bar{\kappa}}) ~{\bar{\mathcal
F}}_{{\bar{\kappa}/\theta_{0i}}}\;, \label{kolejnycocyclefnext}
\end{equation}
respectively\footnote{${{\mathcal F}}_{{\theta_{0i},\bar{\kappa}}} =
{\rm exp}\,i(\frac{1}{2\bar{\kappa}}P_i \wedge M_{kl} +
{{\theta}_{0i}}P_{0 }\wedge P_{i})$.}.  Then, we have
\begin{eqnarray}
\Delta_{\theta_{0i},\bar{\kappa}}(P_\mu)&=&\Delta
_0(P_\mu)+\sin(\frac{1}{2\bar\kappa} P_{i} )\wedge
\left(\eta_{k \mu}P_l -\eta_{l \mu}P_k \right)\label{nextnowageneracja1}\\
&+&(\cos(\frac{1}{2\bar\kappa}  P_{i} )-1)\perp \left(\eta_{k
\mu}P_k +\eta_{l \mu}P_l \right)\;,  \notag
\end{eqnarray}
\begin{eqnarray}
\Delta_{\theta_{0i},\bar{\kappa}}(M_{\mu\nu})&=&\Delta_0(M_{\mu\nu})+\frac{1}{2\bar\kappa}M_{k
l }\wedge  \left(\eta_{\mu
i }P_\nu-\eta_{\nu i }P_\mu\right)\notag\\
&+&i\left[M_{\mu\nu},M_{k l }\right]\wedge
\sin(\frac{1}{2\bar\kappa}P_{i} ) \notag \\
&+&\left[\left[%
M_{\mu\nu},M_{k l }\right],M_{k l }\right]\perp
(\cos(\frac{1}{2\bar\kappa} P_{i} )-1)  \label{nextnowageneracja3} \\
&+&\frac{1}{2\bar\kappa}M_{k l }\sin(\frac{1}{2\bar\kappa} P_{i}
)\perp
 \left(\psi_{i} P_k -\chi_{i} P_l \right) \nonumber \\
&+&\frac{1}{2\bar\kappa} \left(\psi_{i} P_l +\chi_{i} P_k
\right)\wedge M_{k l }(\cos(\frac{1}{2\bar\kappa} P_{i}
)-1)\nonumber
\end{eqnarray}
\begin{eqnarray}
&-& \theta_{0i}[(\eta _{0 \mu }P_{\nu }-\eta _{0 \nu }\,P_{\mu
})\otimes P_{i }+P_{0}\otimes (\eta_{i
\mu}P_{\nu}-\eta_{i \nu}P_{\mu})]\nonumber\\
&+& \theta_{0i}[(\eta _{i \mu }P_{\nu }-\eta _{i \nu }\,P_{\mu
})\otimes P_{0 }+P_{i}\otimes (\eta_{0 \mu}P_{\nu}-\eta_{0
\nu}P_{\mu})]\nonumber\\
&+& \theta_{0i}\left[\left[M_{\mu\nu},M_{kl }\right],P_0\right]
\perp
\sin(\frac{1}{2 {\bar\kappa}} P_i )P_i\notag\\
&-& \theta_{0i}\left[\left[M_{\mu\nu},M_{kl }\right],P_i\right]
\perp \sin(\frac{1}{2 {\bar\kappa}} P_i )P_0
\notag\\
&-&i\theta_{0i}\left[\left[\left[%
M_{\mu\nu},M_{kl }\right],M_{kl }\right],P_0\right] \wedge
(\cos(\frac{1}{2 {\bar\kappa}}  P_i  )-1)P_i
\notag\\
&+& i\theta_{0i}\left[\left[\left[%
M_{\mu\nu},M_{kl }\right],M_{kl }\right],P_i\right] \wedge
(\cos(\frac{1}{2 {\bar\kappa}}  P_i )-1)P_0\;.\notag
\end{eqnarray}
 The above relations together with classical algebraic
sector (\ref{nnn}) and undeformed antipodes/counits (\ref{zadruga1})
 define generalized Poincare Hopf algebra
$\,{\mathcal U}_{\theta_{0i},{\bar \kappa}}({\mathcal P})$.
Obviously, for ${\bar \kappa} \to \infty$ we get $\,{\mathcal
U}_{\theta_{0i}}({\mathcal P})$ Poincare Hopf algebra, while for
$\theta_{0i} \to 0$ we obtain $\,{\mathcal U}_{{\bar
\kappa}}({\mathcal P})$ Hopf structure  described in second Section.

\section{{{Generalized twisted relativistic space-times}}}

In this Section we introduce the generalized relativistic
space-times corresponding to the Poincare Hopf algebras provided in
pervious Section. They are defined as  quantum representation spaces
(Hopf modules) for quantum Poincare algebras, with action of the
deformed symmetry generators satisfying suitably deformed Leibnitz
rules \cite{bloch}, \cite{3b}, \cite{chi}. The action of Poincare
algebra  on a Hopf module of functions depending on space-time
coordinates ${x}_\mu$ is given by

\begin{equation}
P_{\mu }\rhd f(x)=i\partial _{\mu }f(x)\;\;\;,\;\;\; M_{\mu \nu
}\rhd f(x)=i\left( x_{\mu }\partial _{\nu }-x_{\nu }\partial _{\mu
}\right) f(x)\;,  \label{a1}
\end{equation}
while the $\star_{.}$-multiplication of arbitrary two functions  is
defined as follows
\begin{equation}
f({x})\star_{{\cdot}} g({x}):= \omega\circ\left(
 \mathcal{F}_{\cdot}^{-1}\rhd  f({x})\otimes g({x})\right) \;.
\label{star}
\end{equation}
In the above formula $\mathcal{F}_{\cdot}$ denotes  twist factor
corresponding to a proper Poincare group and $\omega\circ\left(
a\otimes b\right) = a\cdot b$.

\subsection*{{\bf i)} the deformation (5)}

$~~~~~$Let us start with $\,{\mathcal
U}_{\theta_{kl},{\kappa}}({\mathcal P})$ Hopf algebra described by
matrix (\ref{rge1}). In such a case, in accordance with
considerations of pervious Section, the star multiplication of two
arbitrary functions $f(x)$ and $g(x)$ is given
by
\begin{eqnarray}
f(x) \star_{\theta_{kl},{\kappa}} g(x) = \omega \circ\left ({\cal
F}_{\theta_{kl},{\kappa}}^{-1} \rhd (f(x)\otimes g(x))
\right)\;,\label{multi1}
\end{eqnarray}
where  the  multiplication operator ${\cal
F}_{\theta_{kl},{\kappa}}$ is defined by the  superposition of two
twist factors (\ref{gfactor}) and (\ref{dghfactor})
\begin{eqnarray}
&&{\cal F}_{\theta_{kl},{\kappa}} = {\cal F}_{\theta_{kl}}\cdot
{\cal F}_{{\kappa}} = {\cal F}_{{\kappa}}\cdot {\cal
F}_{\theta_{kl}} = {\rm exp}~ \frac{i}{2} \left(2\theta_{kl}(P_{k
}\wedge P_{l}) + \frac{1}{{\kappa}}(P_k\wedge
M_{i0})\right)\;.\nonumber\label{starop1}
\end{eqnarray}
Hence, due to the formulas (\ref{a1}), (\ref{star}) we
have\footnote{$[\,a,b\,]_{\star} = a\star b - b\star a$.}
\begin{eqnarray}
&&[\,x_0,x_a\,]_{{\star}_{\theta_{kl},{\kappa}}}
=\frac{i}{\kappa}x_i\delta_{ak}\;\;\;,\;\;\;
[\,x_a,x_b\,]_{{\star}_{\theta_{kl},{\kappa}}}
=2i\theta_{kl}(\delta_{ak}\delta_{bl} - \delta_{al}\delta_{bk}) +
\frac{i}{\kappa}x_0 (\delta_{ia}\delta_{kb} -
\delta_{ka}\delta_{ib}) \;.\nonumber\label{spacetime1}
\end{eqnarray}
Of course, for parameter $\theta_{kl}$ approaching zero or parameter
$\kappa$ running to infinity, we get the relativistic space-times
associated with Hopf algebras $\,{\mathcal U}_{{\kappa}}({\mathcal
P})$ or $\,{\mathcal U}_{\theta_{kl}}({\mathcal P})$ respectively
(see \cite{chi}, \cite{lie2}).

\subsection*{{\bf ii)} the deformation (6)}

$~~~~~$In the case of $\,{\mathcal U}_{\theta_{0i},{\hat
\kappa}}({\mathcal P})$ Hopf algebra the corresponding
multiplication looks as follows
\begin{eqnarray}
f(x) \star_{\theta_{0i},{\hat \kappa}} g(x) = \omega \circ\left
({\cal F}_{\theta_{0i},{\hat \kappa}}^{-1} \rhd (f(x)\otimes g(x))
\right)\;,\label{multi2}
\end{eqnarray}
with operator ${\cal F}_{\theta_{0i},{\hat \kappa}}$ given by
\begin{eqnarray}
&&{\cal F}_{\theta_{0i},{\hat \kappa}} = {\cal F}_{\theta_{0i}}\cdot
{\cal F}_{\hat {\kappa}} = {\cal F}_{{\hat \kappa}}\cdot {\cal
F}_{\theta_{0i}} = {\rm exp}~ \frac{i}{2} \left(2\theta_{0i}(P_{0
}\wedge P_{i}) + \frac{1}{\hat {\kappa}}(P_0\wedge
M_{kl})\right)\;.\nonumber\label{starop2}
\end{eqnarray}
Consequently, we get
\begin{eqnarray}
[\,x_0,x_a\,]_{{\star}_{\theta_{0i},{\hat \kappa}}} =\frac{i}{{\hat
\kappa}}(\delta_{la}x_k - \delta_{ka}x_l) + 2i\theta_{0i}\delta_{ia}
\;\;\;,\;\;\; [\,x_a,x_b\,]_{{\star}_{\theta_{0i},{\hat \kappa}}} =0
\;.\label{spacetime2}
\end{eqnarray}
The above relations define the relativistic space-time corresponding
to the algebra $\,{\mathcal U}_{\theta_{0i},{\hat \kappa}}({\mathcal
P})$. Obviously, for $\hat {\kappa} \to \infty$ or $\theta_{0i} \to
0$ we obtain the space-time associated with $\,{\mathcal
U}_{\theta_{0i}}({\mathcal P})$ or $\,{\mathcal U}_{{\hat
\kappa}}({\mathcal P})$ Hopf algebra provided in \cite{chi} and
\cite{lie2} respectively.

\subsection*{{\bf iii)} the deformation (7)}

$~~~~~$For the last Hopf algebra $\,{\mathcal U}_{\theta_{0i},{\bar
\kappa}}({\mathcal P})$ we have
\begin{eqnarray}
f(x) \star_{\theta_{0i},{\bar \kappa}} g(x) = \omega \circ\left
({\cal F}_{\theta_{0i},{\bar \kappa}}^{-1} \rhd (f(x)\otimes g(x))
\right)\;,\label{multi3}
\end{eqnarray}
and
\begin{eqnarray}
{\cal F}_{\theta_{0i},{\bar \kappa}} = {\rm exp}~ {\frac{i}{2}
\left(2\theta_{0i}(P_{0 }\wedge P_{i}) + \frac{1}{\bar {\kappa}}(P_i
\wedge M_{kl})\right)}\;, \label{multi3df}
\end{eqnarray}
what gives the following relativistic space-time
\begin{eqnarray}
&&[\,x_0,x_a\,]_{{\star}_{\theta_{0i},{\bar \kappa}}}
=2i\theta_{0i}\delta_{ia}\;\;\;,\;\;\;
[\,x_a,x_b\,]_{{\star}_{\theta_{0i},{\bar \kappa}}} = \frac{i}{{\bar
\kappa}}\delta_{ib}(\delta_{ka}x_l - \delta_{la}x_k) +
\frac{i}{{\bar \kappa}}\delta_{ia}(\delta_{lb}x_k - \delta_{kb}x_l)
\;.\nonumber\label{spacetime3}
\end{eqnarray}
Of course, for parameter ${\bar \kappa}$ approaching infinity or
parameter $\theta_{0i}$ running to zero, we get the relativistic
space-time corresponding to $\,{\mathcal U}_{\theta_{0i}}({\mathcal
P})$ or $\,{\mathcal U}_{{\bar \kappa}}({\mathcal P})$ Poincare Hopf
algebras respectively.

\section{{{Generalized twisted Galilei Hopf algebras}}}

In this Section we calculate the  nonrelativistic contractions of
Hopf structures derived in Section 3, i.e. we find their
nonrelativistic counterparts - the generalized twist deformations of
Galilei Hopf algebra.

First of all, let us introduce the following  redefinition of
Poincare generators \cite{inonu}
\begin{equation}
P_{0 } = \frac{{\Pi_{0 }}}{c}\;\;\;,\;\;\;P_{i } =
\Pi_{i}\;\;\;,\;\;\;M_{ij}= K_{ij}\;\;\;,\;\;\;M_{i0}= cV_i\;,
\label{contr2}
\end{equation}
where parameter $c$ denotes the light velocity. Besides,  we also
introduce five  parameters $\lambda$, $\hat{\lambda}$,
$\overline{\lambda}$, $\xi_{kl}$ ($\xi_{lk}$) and $\xi_{0i}$
($\xi_{i0}$) such that
\begin{equation}
\lambda = {\kappa}/{c}\;\;,\;\; \hat{\lambda} = \hat{\kappa}
c\;\;,\;\;\overline{\lambda} = \overline{\kappa}\;\;,\;\;\xi_{kl} =
\theta_{kl}\;\;(\xi_{lk}=\theta_{lk})\;\;,\;\;\xi_{0i} =
{\theta_{0i}}/{c}\;\;(\xi_{i0} = {\theta_{i0}}/{c}) \;.
\label{contr3}
\end{equation}
 Further,
one performs the contraction limit of
 algebraic part (\ref{nnn}) and coproducts
 ((\ref{coa1}),(\ref{coa100})), ((\ref{nextczartworfff}),(\ref{nextczartworcopp2})) and
 ((\ref{nextnowageneracja1}),(\ref{nextnowageneracja3})) in two
 steps. Firstly, we rewrite the formulas (\ref{nnn}) and ((\ref{coa1}),(\ref{coa100})),
 ((\ref{nextczartworfff}),(\ref{nextczartworcopp2})),
 ((\ref{nextnowageneracja1}),(\ref{nextnowageneracja3}))
 in term of the operators
(\ref{contr2}) and parameters (\ref{contr3}). Secondly, we take the
$c\to \infty$ limit, and in such a way, we get the following
algebraic\footnote{$a, b, c, d =1, 2,3$.}
\begin{eqnarray}
&&\left[\, K_{ab},K_{cd}\,\right] =i\left( \delta
_{ad}\,K_{bc}-\delta
_{bd}\,K_{ac}+\delta _{bc}K_{ad}-\delta _{ac}K_{bd}\right) \;,  \notag \\
&&\left[\, K_{ab},V_{c}\,\right] =i\left( \delta _{bc}\,V_a-\delta
_{ac}\,V_b\right)\;\; \;, \;\;\;\left[ \,K_{ab},\Pi_{c }\,\right]
=i\left( \delta _{b c }\,\Pi_{a }-\delta _{ac }\,\Pi_{b }\right) \;,
\label{nnnga}
\\
&&\left[ \,K_{ab},\Pi_{0 }\,\right] =\left[ \,V_a,V_b\,\right] =
\left[ \,V_a,\Pi_{b }\,\right] =0\;\;\;,\;\;\;\left[ \,V_a,\Pi_{0
}\,\right] =-i\Pi_a\;\;\;,\;\;\;\left[ \,\Pi_{\rho },\Pi_{\sigma
}\,\right] = 0\;,\nonumber
\end{eqnarray}
and coalgebraic sectors

\subsection*{{\bf i)} nonrelativistic $\bf(\xi_{kl},\lambda)$-deformation}

\begin{eqnarray}
 \Delta_{\xi_{kl},{\lambda}}(\Pi_0)&=&\Delta _0(\Pi_0) +
\frac{1}{2{{\lambda}}} \Pi_k \wedge \Pi_i\;,\label{gacoa1}\\
\Delta_{\xi_{kl},{\lambda}}(\Pi_a)&=&\Delta
_0(\Pi_a)\;\;\;,\;\;\;\Delta_{\xi_{kl},{\lambda}}(V_a)=\Delta_0(V_a)\;,\label{coa0}\\
 &~~&  \cr
\Delta_{\xi_{kl},{\lambda}}(K_{ab})&=&\Delta_0(K_{ab})+
\frac{i}{2{{\lambda}}}\left[K_{ab},V_i\right]\wedge \Pi_k +
\frac{1}{2{{\lambda}}}V_i \wedge(\delta_{ak}\Pi_b
-\delta_{bk}\Pi_a) \nonumber\\ 
&-&\xi_{k l }[(\delta_{k a}\Pi_{b }-\delta_{k b }\,\Pi_{a})\otimes
\Pi_{l }+\Pi_{k}\otimes (\delta_{l a}\Pi_{b}-\delta_{l b}\Pi_{a})]
 \label{gacoa100}\\
 &+&\xi_{k l }[(\delta_{l a}\Pi_{b }-\delta_{l b }\,\Pi_{a})\otimes
\Pi_{k }+\Pi_{l}\otimes (\delta_{ka}\Pi_{b}-\delta_{kb}\Pi_{a})]
\nonumber\;,
\end{eqnarray}

\subsection*{{\bf ii)} nonrelativistic $\bf (\xi_{0i},\hat{\lambda})$-deformation}

\begin{eqnarray}
\Delta_{\xi_{0i},\hat{\lambda}}(\Pi_\mu)&=&\Delta _0(\Pi_\mu)+\sin(
\frac{1}{2{\hat \lambda}} \Pi_0 )\wedge
\left(\delta_{k \mu}\Pi_l -\delta_{l \mu}\Pi_k \right)\nonumber\\
&+&(\cos(\frac{1}{2{\hat \lambda}}  \Pi_0 )-1)\perp \left(\delta_{k
\mu}\Pi_k +\delta_{l \mu}\Pi_l \right)\;, \label{ganextczartworfff}
\end{eqnarray}
\begin{eqnarray}
\Delta_{\xi_{0i},\hat{\lambda}}(K_{ab})&=&\Delta_0(K_{ab})+i\left[K_{ab},K_{k
l }\right]\wedge \sin(\frac{1}{2{\hat \lambda}} \Pi_0 )\notag\\
&+&\left[\left[%
K_{ab},K_{k l }\right],K_{k l }\right]\perp (\cos(\frac{1}{2{\hat
\lambda}}  \Pi_0 )-1)\;\label{ganextczartworfff100} \\
&-& \xi_{0i}\Pi_0 \wedge\left(\delta_{ia}\Pi_b -
\delta_{ib}\Pi_a\right)\;\notag\\
&-&\xi_{0i}\left[\left[K_{ab},K_{k l }\right],\Pi_i\right]
\perp \Pi_0 \sin(\frac{1}{2{\hat \lambda}} \Pi_0 )\notag\\
&+&i\xi_{0i}\left[\left[\left[%
K_{ab},K_{k l }\right],K_{k l }\right],\Pi_i\right]\wedge
\Pi_0(\cos(\frac{1}{2{\hat \lambda}}  \Pi_0 )-1)\;,\notag
\end{eqnarray}
\begin{eqnarray}
\Delta_{\xi_{0i},\hat{\lambda}}(V_a)&=&\Delta_0(V_a)+\frac{1}{2{\hat{\lambda}}}K_{k
l}\wedge
 \Pi_a+i\left[V_a,K_{k l }\right]\wedge
\sin(\frac{1}{2{\hat{\lambda}}} \Pi_0 ) \notag \\
&+&\left[\left[%
V_a,K_{k l }\right],K_{k l }\right]\perp
(\cos(\frac{1}{2{\hat{\lambda}}}  \Pi_0  )-1)
\notag\\
&+&K_{k l }\sin(\frac{1}{2{\hat{\lambda}}} \Pi_0 )\perp
\frac{1}{2{\hat{\lambda}}} \left(\delta_{k a}\Pi_l - \delta_{l a} \Pi_k  \right) \notag \\
&-&\frac{1}{2{\hat{\lambda}}} \left(\delta_{k a}\Pi_k+\delta_{l a}
\Pi_l \right)\wedge K_{k l }(\cos(\frac{1}{2{\hat{\lambda}}} \Pi_0
)-1)
  \label{ganextczartworcopp2}\\
&-& {\xi_{0i}}\Pi_a \wedge \Pi_i -i\xi_{0i}\left[\left[\left[%
V_a,K_{k l }\right],K_{k l }\right],\Pi_0\right]\wedge
(\cos(\frac{1}{2{\hat{\lambda}}} \Pi_0  )-1)\Pi_i \notag\\
&+&\xi_{0i}\left[\left[V_a,K_{k l
}\right],\Pi_0\right]\perp\sin(\frac{1}{2{\hat{\lambda}}} \Pi_0
)\Pi_i\;,\notag
\end{eqnarray}

\subsection*{{\bf iii)} nonrelativistic $\bf (\xi_{0i},\bar{\lambda})$-deformation}

\begin{eqnarray}
\Delta_{\xi_{0i},\bar \lambda}(\Pi_\mu)&=&\Delta _0(\Pi_\mu)+\sin(
\frac{1}{2{\bar \lambda}} \Pi_i )\wedge
\left(\delta_{k \mu}\Pi_l -\delta_{l \mu}\Pi_k \right)\nonumber\\
&+&(\cos(\frac{1}{2{\bar \lambda}}  \Pi_i )-1)\perp \left(\delta_{k
\mu}\Pi_k +\delta_{l \mu}\Pi_l \right)\;,
\label{genextnowageneracja1}
\end{eqnarray}
\begin{eqnarray}
~~ \Delta_{\xi_{0i},\bar \lambda}(K_{ab})&=&\Delta_0(K_{ab})+K_{k l
}\wedge \frac{1}{2{\bar \lambda}} \left(\delta_{a
i }\Pi_b-\delta_{b i }\Pi_a\right)\nonumber\\
&+&i\left[K_{ab},K_{k l }\right]\wedge
\sin(\frac{1}{2{\bar \lambda}} \Pi_i ) \notag \\
&+&\left[\left[%
K_{ab},K_{k l }\right],K_{k l }\right]\perp
(\cos(\frac{1}{2{\bar \lambda}}  \Pi_i  )-1) \notag \\
&+&K_{k l }\sin(\frac{1}{2{\bar \lambda}} \Pi_i )\perp
\frac{1}{2{\bar \lambda}} \left(\psi_i \Pi_k -\chi_i \Pi_l \right) \label{genextnowageneracja2}  \\
&+&\frac{1}{2{\bar \lambda}} \left(\psi_i \Pi_l +\chi_i \Pi_k
\right)\wedge K_{k l }(\cos(\frac{1}{2{\bar \lambda}} \Pi_i )-1)
  \notag
\end{eqnarray}
\begin{eqnarray}
&-& \xi_{0i}\Pi_0 \wedge\left(\delta_{ia}\Pi_b -
\delta_{ib}\Pi_a\right)\;\notag\\
&-&\xi_{0i}\left[\left[K_{ab},K_{k l }\right],\Pi_i\right]
\perp \Pi_0 \sin(\frac{1}{2{\bar \lambda}} \Pi_i )\notag\\
&+&i\xi_{0i}\left[\left[\left[%
K_{ab},K_{k l }\right],K_{k l }\right],\Pi_i\right]\wedge
\Pi_0(\cos(\frac{1}{2{\bar \lambda}}  \Pi_i )-1)\;,\notag
\end{eqnarray}
\begin{eqnarray}
\Delta_{\xi_{0i},\bar \lambda}(V_a)&=&\Delta_0(V_a) +i\left[V_a,K_{k
l }\right]\wedge \sin(\frac{1}{2{\bar \lambda}} \Pi_i )
\label{genextnowageneracja3}\\
&+&\left[\left[%
V_a,K_{k l }\right],K_{k l }\right]\perp (\cos(\frac{1}{2{\bar
\lambda}} \Pi_i )-1) \nonumber\\
&-& {\xi_{0i}}\Pi_a \wedge \Pi_i -i\xi_{0i}\left[\left[\left[%
V_a,K_{k l }\right],K_{k l }\right],\Pi_0\right]\wedge
(\cos(\frac{1}{2{\bar{\lambda}}} \Pi_i  )-1)\Pi_i \notag\\
&+&\xi_{0i}\left[\left[V_a,K_{k l
}\right],\Pi_0\right]\perp\sin(\frac{1}{2{\bar{\lambda}}} \Pi_i
)\Pi_i\;.\notag
\end{eqnarray}
The relations (\ref{nnnga}) together with coproducts ${\bf i)}$,
${\bf ii)}$ and ${\bf iii)}$ define the generalized twisted Galilei
Hopf algebras $\,{\mathcal U}_{\xi_{kl},{\lambda}}({\mathcal G})$,
$\,{\mathcal U}_{\xi_{0i},{\hat \lambda}}({\mathcal G})$ and
$\,{\mathcal U}_{\xi_{0i},{\bar \lambda}}({\mathcal G})$
corresponding to the Poincare Hopf algebras $\,{\mathcal
U}_{\theta_{kl},{\kappa}}({\mathcal P})$, $\,{\mathcal
U}_{\theta_{0i},{\hat \kappa}}({\mathcal P})$ and $\,{\mathcal
U}_{\theta_{0i},{\bar \kappa}}({\mathcal P})$ respectively. It
should be noted that for parameters $\xi_{kl}$, $\xi_{0i}$ running
to zero and (or) parameters $\lambda$, ${\hat \lambda}$, ${\bar
\lambda}$ approaching infinity, the   above algebras become
classical (or one gets twisted Galilei Hopf structures proposed in
\cite{dasz}).

\section{{{Final remarks}}}

In this article we provide three new generalized Poincare Hopf
algebras and corresponding relativistic space-times. All three
space-times combine two kinds of quantum deformations - canonical
and Lie-algebraic deformation  leading to the models  with quantum
time and classical space as well as  with quantum time and quantum
space. Further, we also perform three nonrelativistic contraction
limits to the corresponding Galilei Hopf structures.

The present studies can be  extended in various ways. First of all,
one can ask about $N=1$ supersymmetric extensions of the constructed
deformed Hopf  algebras. Besides, 
one can also consider still more complicated (twisted)
generalizations of the relativistic and nonrelativistic quantum
space-times. For example, it is possible to consider the Poincare or
Galilei Hopf structure leading  to the superposition of canonically
and quadratically deformed quantum space. Finally, it should be
noted, that  dual quantum Poincare (Galilei) groups $\,{\mathcal
P}_{\theta_{kl},{\kappa}}$ $\,({\mathcal G}_{\xi_{kl},{\lambda}})$,
$\,{\mathcal P}_{\theta_{0i},{\hat \kappa}}$ $\,({\mathcal
G}_{\xi_{0i},{\hat \lambda}})$ and $\,{\mathcal
P}_{\theta_{0i},{\bar \kappa}}$ $\,({\mathcal G}_{\xi_{0i},{\bar
\lambda}})$ can be obtained by canonical quantization of the
corresponding Poisson-Lie structure \cite{poisson} or with use of
so-called FRT procedure \cite{frt}. Consequently, as it was
mentioned in Introduction, one can find in the framework of
Heisenberg double procedure \cite{twist1} the corresponding
relativistic and nonrelativistic phase spaces associated with the
above algebras. All these problems are now being studied.

\section*{Acknowledgments}
The author would like to thank J. Lukierski and M. Woronowicz
for valuable discussions.\\
This paper has been financially supported by Polish Ministry of
Science and Higher Education grant NN202318534.

\end{document}